\documentclass[aps,pre,twocolumn,showpacs]{revtex4}
\usepackage{graphicx}

\begin{document}

\title{Alternative interpretation of the sign reversal of secondary Bjerknes 
force acting between two pulsating gas bubbles}
\author{Masato Ida}
\affiliation{Collaborative Research Center of Frontier Simulation Software for Industrial Science, 
Institute of Industrial Science, the University of Tokyo, 
4--6--1 Komaba, Meguro-Ku, Tokyo 153--8505, Japan
}

\begin{abstract}
It is known that in a certain case, the secondary Bjerknes force, which is a 
radiation force acting between pulsating bubbles, changes, e.g., from 
attraction to repulsion as the bubbles approach each other. In this paper, a 
theoretical discussion of this phenomenon for two spherical bubbles is 
described. The present theory based on analysis of the transition 
frequencies of interacting bubbles [M. Ida, Phys. Lett. A \textbf{297}, 210 
(2002)] provides an interpretation, different from previous ones (e.g., by 
Doinikov and Zavtrak [Phys. Fluids \textbf{7}, 1923 (1995)]), of the 
phenomenon. It is shown, for example, that the reversal that occurs when one 
bubble is smaller and another is larger than a resonance size is due to the 
second-highest transition frequency of the smaller bubble, which cannot be 
obtained using traditional natural-frequency analysis.
\end{abstract}

\maketitle

\section{INTRODUCTION}
\label{sec1}
It is known that two gas bubbles pulsating in an acoustic field undergo an 
interaction force called the secondary Bjerknes force \cite{ref1,ref2,ref3}. This 
force is attractive when the bubbles pulsate in-phase with each other, while 
it is repulsive otherwise; that is, the phase property of the bubbles plays 
an important role in determining the sign of the force. In a seminal paper 
published in 1984 \cite{ref4}, Zabolotskaya, using a linear coupled oscillator 
model, showed theoretically that in a certain case, the sign of the force 
may change as the bubbles come closer to one another. This theoretical 
prediction was ensured by recent experiments that captured the stable, 
periodic translational motion of two coupled bubbles \cite{ref12}, resulting from 
the sign reversal of the force at a certain distance between the bubbles. 
Zabolotskaya assumed that this sign reversal is due to variation in the 
natural frequencies of the interacting bubbles, which results in shifts of 
their pulsation phases. The theoretical formula Zabolotskaya derived to 
evaluate the natural frequencies of two interacting bubbles, which 
corresponds to one given previously by Shima \cite{ref5}, is represented as
\begin{equation}
\label{eq1}
(\omega _{10}^2 - \omega ^2)(\omega _{20}^2 - \omega ^2) - \frac{R_{10} 
R_{20} }{D^2}\omega ^4 \approx 0,
\end{equation}
where $R_{10} $ and $R_{20} $ are the equilibrium radii of the bubbles, 
$\omega _{10} $ and $\omega _{20} $ are their partial natural (angular) 
frequencies, $\omega $ is the angular frequency of an external sound, and 
$D$ is the distance between the centers of the bubbles. This equation 
predicts the existence of two natural frequencies per bubble, and is 
symmetric; namely, it exchanges 10 and 20 in the subscripts of the variables 
to reproduce the same equation, meaning that the two bubbles have the same 
natural frequencies.

During the last decade, a number of studies regarding the sign reversal of 
the force have been performed \cite{ref6,ref7,ref8,ref9,ref10,ref11,ref12,ref13,ref20}. Among them, Refs.~\cite{ref8,ref9,ref13} also considered the relevance of the change in the natural frequencies 
(or resonance frequencies \cite{ref22}) to the sign reversal. In the present paper, 
we focus our attention on this case, although it has been shown that other 
factors, such as the nonlinearity in bubble pulsation \cite{ref6,ref10,ref11,ref13,ref20} 
and the higher-order terms appearing in the time-averaged interaction force 
\cite{ref12} which has been neglected in previous works, can also cause the sign 
reversal.

In 1995, Doinikov and Zavtrak \cite{ref8}, using a linear mathematical model in 
which the multiple scattering of sound between bubbles is taken into account 
more rigorously, predicted again the sign reversal. They also asserted that 
this reversal is due to the change in the natural frequencies. They assumed 
that the natural frequencies of both bubbles increase as the bubbles 
approach each other, resulting sometimes in the sign reversal. When, for 
example, both bubbles are larger than the resonance size (i.e., the case of 
$\omega _{10}  < \omega $ and $\omega _{20}  < \omega $) and the distance 
between them is large enough, they pulsate in-phase with each other. As the 
bubbles approach each other, the natural frequency of a smaller bubble may 
first, at a certain distance, rise above the driving frequency, and in turn 
the bubbles' pulsations become antiphase; the force then changes from 
attractive to repulsive. When, on the other hand, one bubble is larger and 
the other is smaller than the resonance size (e.g., $\omega _{10}  > \omega  > \omega _{20} $) and the distance between them is 
large, they pulsate out-of-phase with each other and the force is repulsive. 
As the distance between the bubbles becomes smaller, the natural frequencies 
of both bubbles may rise, and when the natural frequency of a larger bubble 
rises above the driving frequency, the repulsive force may turn into 
attraction. This interpretation is supported even in more recent papers \cite{ref11,ref20}.

Although this interpretation seems to explain the sign reversal well, it is 
opposed to the prediction given by Eq.~(\ref{eq1}) which reveals that the higher 
natural frequency (converging to the partial natural frequency of a smaller 
bubble for $D \to \infty $ \cite{ref5,ref14}) 
increases but the lower one (converging to the partial natural frequency of 
a larger bubble for $D \to \infty $) decreases as the bubbles approach each other.

In 2001, Harkin \textit{et al.} \cite{ref13} performed an extensive theoretical study concerning 
the translational motion of two acoustically coupled gas bubbles in a weak 
and a moderate driving sound field. Their theoretical model derived from 
first principles supports the experimental results by Barbat \textit{et al.} \cite{ref12}. In Sec.~7 of that paper, Harkin \textit{et al.} also considered the influence of the change in 
natural frequencies on the sign of the force in order to explain the sign 
reversal for $\omega _{10}  < \omega$ and $\omega _{20}  < \omega $. Their explanation 
based on a formula given directly by Eq.~(\ref{eq1}) is essentially the same as 
those by Zabolotskaya \cite{ref4} and by Doinikov {\&} Zavtrak \cite{ref8,ref9}.

The authors should note here that all the previous theoretical models 
mentioned above can \textit{describe} (or \textit{explain}) the sign reversal. However, the \textit{interpretation} we will provide 
in the present paper is different from the previous ones.

The aim of this paper is to give an alternative interpretation of the sign 
reversal, one that may be more accurate than the previous ones that are 
based on the natural-frequency analysis. Recently, having reexamined the 
linear coupled oscillator model used frequently to analyze the dynamics of 
acoustically coupled bubbles (see Ref.~\cite{ref14} and references therein), we 
found that a bubble interacting with a neighboring bubble has three 
``transition frequencies'', defined as \textit{the driving frequencies for which the phase difference between an external sound and the bubble's pulsation becomes} $\pi / {\it 2}$ \textit{(or} ${\it 3} \pi / {\it 2}$\textit{)}, two of which correspond to the natural frequencies \cite{ref14}. Among 
the three transition frequencies, the lowest one decreases and the remaining 
two increase as the bubbles approach each other. Meanwhile, for $D \to \infty$ only one of them 
converges to the partial natural frequency of the corresponding bubble. 
Namely, the transition frequencies defined as above are asymmetric. The use 
of the transition frequencies would allow us an accurate understanding of 
the sign reversal, because observing these frequencies provides more 
detailed insights of the bubbles' phase properties rather than that provided 
by the natural-frequency analysis. Using the theory for the transition 
frequencies, we arrive at a novel interpretation of the sign reversal.

\section{THEORIES}
In this section, we briefly review the previously expounded theories regarding 
the natural frequencies, the transition frequencies, and the secondary 
Bjerknes force.

\subsection{Natural frequencies and transition frequencies}
Let us consider the linear volume oscillation of $N$--bubble system immersed in an 
incompressible liquid. Suppose that the time-dependent radius of bubble 
$j$, $R_j$, can be represented as 
$R_j  = R_{j0}  + e_j (t)$ and $\left| {e_j } \right| \ll R_{j0}$, where 
$R_{j0}$ and $e_j$ are the equilibrium 
radius and the deviation of the radius, respectively, and $j = 1,2, \ldots ,N$
. The radius deviation can be 
determined by solving the linear oscillator model (see, e.g., Ref.~\cite{ref23}),
\begin{equation}
\label{eq2}
\ddot e_j  + \omega _{j0}^2 e_j  + \delta _j \dot e_j  =  - \frac{{p_{{\rm d},j} }}{{\rho R_{j0} }},
\end{equation}
where
\[
\omega _{j0}  = \sqrt {\frac{{3\kappa _j P_0  + (3\kappa _j  - 1)2\sigma /R_{j0} }}{{\rho R_{j0}^2 }}}
\]
are the partial natural (angular) frequencies of bubble $j$, $\delta _j$ is the damping coefficient determined based on the 
damping characteristics of the bubbles \cite{ref15}, $p_{{\rm d},j}$ is the driving pressure acting on bubble $j$, $\kappa _j$ is the effective polytropic exponent 
of the gas inside the bubbles, $P_0$ is the static pressure, $\sigma$ is the surface tension, $\rho$ is the density of 
the liquid surrounding the bubbles, and the overdots denote the time 
derivation. The driving pressure is represented by the sum of 
$p_{{\rm ex}}$ and the sound 
pressure scattered by the surrounding bubbles, $p_{\rm s}$, as
\[
p_{{\rm d},j}  = p_{{\rm ex}}  + \sum\limits_{k = 1,k \ne j}^N {p_{{\rm s},j\,k} } .
\]
The value of $p_{{\rm s},j\,k}$ is determined by integrating the momentum equation for linear sound waves, 
$\partial p/\partial r =  - \rho \partial u/\partial t$, coupled with the 
divergence-free condition, $\partial (r^2 u)/\partial r = 0$, where $r$ is 
the radial coordinate measured from the center of a bubble and 
$u$ is the velocity along $r$. Resultantly, the driving pressure is determined as
\begin{equation}
\label{eq3}
p_{{\rm d},j}  = p_{{\rm ex}}  + \rho \sum\limits_{k = 1,k \ne j}^N {\frac{{R_{k0}^2 }}{{D_{j\,k} }}\ddot e_k } ,
\end{equation}
where $D_{j\,k}$ is the distance between the centers of bubbles $j$ and $k$.

In a single-bubble case (i.e., for $N = 1$), Eq.~(\ref{eq2}) is reduced to
\begin{equation}
\label{eq4}
\ddot e_1  + \omega _{10}^2 e_1  + \delta _1 \dot e_1  =  - \frac{{p_{{\rm ex}} }}{{\rho R_{10} }} .
\end{equation}
Assuming that $p_{{\rm ex}}$ is written in the form of $p_{{\rm ex}}  =  - P_a \sin \omega t$ ($P_a$ is a 
positive constant), the harmonic steady-state solution of Eq.~(\ref{eq4}) is given 
by
\[
e_1  = K_{S1} \sin (\omega t - \phi _{S1} ) ,
\]
with
\begin{eqnarray*}
&&K_{S1}  = \frac{{P_a }}{{\rho _0 R_{10} }}\sqrt {\frac{1}{{(\omega _{10}^2  - \omega ^2 )^2  + \delta _1^2 \omega ^2 }}} , \\
&&\phi _{S1}  = \tan ^{ - 1} \left( {\frac{{\delta _1 \omega }}{{\omega _{10}^2  - \omega ^2 }}} \right) .
\end{eqnarray*}
From this result, one knows that the phase difference of $\phi _{S1}  = \pi /2$
 appears (or, roughly speaking, the 
phase reversal takes place) only at the natural frequency $\omega _{10}$ \cite{ref24}, and the resonance response occurs at (or, more correctly, near) the same driving frequency.

For $N = 2$, Eq.~(\ref{eq2}) is reduced to
\begin{eqnarray}
\label{eq5}
&&\ddot e_1  + \omega _{10}^2 e_1  + \delta _1 \dot e_1  =  - \frac{{p_{{\rm ex}} }}{{\rho R_{10} }} - \frac{{R_{20}^2 }}{{R_{10} D}}\ddot e_2 , \\
\label{eq6}
&&\ddot e_2  + \omega _{20}^2 e_2  + \delta _2 \dot e_2  =  - \frac{{p_{{\rm ex}} }}{{\rho R_{20} }} - \frac{{R_{10}^2 }}{{R_{20} D}}\ddot e_1 ,
\end{eqnarray}
where $D = D_{12}  = D_{21}$. It is known 
that for a weak forcing (i.e., $P_a  \ll P_0$), this system has third-order accuracy with respect to $1/D$, although it has 
terms of up to first order (the last terms) \cite{ref13}. The harmonic steady-state solution for $e_1$ is
\[
e_1  = K_1 \sin (\omega t - \phi _1 ) ,
\]
where
\begin{eqnarray*}
&&K_1  = \frac{{P_a }}{{R_{10} \rho }}\sqrt {A_1^2  + B_1^2 } , \\
&&\phi _1  = \tan ^{ - 1} \left( {\frac{{B_1 }}{{A_1 }}} \right) \in [0,2\pi ] ,
\end{eqnarray*}
with
\begin{eqnarray*}
&&A_1  = \frac{{H_1 F + M_2 G}}{{F^2  + G^2 }},
\quad
B_1  = \frac{{H_1 G - M_2 F}}{{F^2  + G^2 }}, \\
&&F = L_1 L_2  - \frac{{R_{10} R_{20} }}{{D^2 }}\omega ^4  - M_1 M_2 , \\
&&G = L_1 M_2  + L_2 M_1 ,
\quad
H_1  = L_2  + \frac{{R_{20} }}{D}\omega ^2 , \\
&&L_1  = \omega _{10}^2  - \omega ^2 ,
\quad
L_2  = \omega _{20}^2  - \omega ^2 , \\
&&M_1  = \delta _1 \omega ,
\quad
M_2  = \delta _2 \omega .
\end{eqnarray*}
Exchanging 1 and 2 (or 10 and 20) in the subscripts of these equations 
yields the expressions for bubble 2.

The formula for the natural frequency, Eq.~(\ref{eq1}), is derived so that 
$K_1  \to \infty $ for $\delta _1  \to 0$ and $\delta _2  \to 0$. Namely,
\[
F = L_1 L_2  - \frac{{R_{10} R_{20} }}{{D^2 }}\omega ^4  = 0 .
\]
As mentioned already, this equation predicts the existence of up to two 
natural frequencies in a double-bubble system.

The transition frequencies of bubble 1 are determined so that 
$\phi _1$ becomes $\pi /2$ (or $3\pi /2$). Because 
$F^2  + G^2  \ne 0$ \cite{ref14}, the resulting 
formula for deriving the transition frequencies of bubble 1 is
\begin{equation}
\label{eq7}
H_1 F + M_2 G = 0 .
\end{equation}
Assuming $\delta _1  \to 0$ and $\delta _2  \to 0$ reduces this to
\begin{equation}
\label{eq8}
H_1 F = \left( {L_2  + \frac{{R_{20} }}{D}\omega ^2 } \right)\left( {L_1 L_2  - \frac{{R_{10} R_{20} }}{{D^2 }}\omega ^4 } \right) = 0 .
\end{equation}
As was proven in Ref.~\cite{ref14}, this equation predicts the existence of up to 
three transition frequencies per bubble. Furthermore, as pointed out in the 
same article, the terms in the second $\left(  \cdots  \right)$ of Eq.~(\ref{eq8}) are the same as those on the left-hand 
side of Eq.~(\ref{eq1}). These results mean that in a double-bubble case the phase 
reversal of a bubble's pulsation can take place not only at its natural 
frequencies but also at one other frequency. Because $H_1  \ne H_2 $, Eq.~(\ref{eq8}) (and also Eq.~(\ref{eq7})) is 
asymmetric, meaning that the bubbles have different transition frequencies.

A preliminary discussion for a $N$--bubble system \cite{ref19} showed that a bubble in the 
system has up to $2N - 1$ transition frequencies, $N$ ones of which correspond to the natural frequency. Namely, a 
bubble has an odd number of transition frequencies. This result can be 
understood as follows: Even in a multibubble case, a bubble's pulsation may 
be in-phase or out-of-phase with a driving sound \cite{ref25} when the driving 
frequency is much lower or much higher, respectively, than its natural 
frequencies; thus, in order to interpolate these two extremes consistently 
an odd number of phase reversals are necessary \cite{ref19}.

\subsection{Secondary Bjerknes force}
The secondary Bjerknes force acting between the bubbles for sufficiently 
weak forcing is expressed with \cite{ref1,ref2,ref3,ref4,ref12,ref13}
\begin{equation}
\label{eq9}
{\bf F} \propto \left\langle {\dot V_1 \dot V_2 } \right\rangle \frac{{{\bf r}_2  - {\bf r}_1 }}{{\left\| {{\bf r}_2  - {\bf r}_1 } \right\|^3 }} \propto K_1 K_2 \cos (\phi _1  - \phi _2 )\frac{{{\bf r}_2  - {\bf r}_1 }}{{\left\| {{\bf r}_2  - {\bf r}_1 } \right\|^3 }} ,
\end{equation}
where $V_j$ and ${\bf r}_j$ are the volume and 
the position, respectively, of bubble $j$, $\left\langle \cdots \right\rangle $ denotes the time average, and $\left\| {{\bf r}_2  - {\bf r}_1 } \right\| = D$. The sign reversal of this force occurs only when 
the sign of $\cos (\phi _1  - \phi _2 )$ (or of $\left\langle {\dot V_1 \dot V_2 } \right\rangle $) changes, because $K_1  > 0$ and $K_2  > 0$. If the phase 
shifts resulting from the radiative interaction between bubbles are 
neglected, this force is repulsive when $\omega$ stays between $\omega _{10}$ and $\omega _{20}$, and is attractive otherwise \cite{ref1}. In the case where the radiative 
interaction is taken into consideration, the frequency within which the 
force is repulsive shifts toward a higher range, see, e.g., Refs.~\cite{ref8,ref9}.

\subsection*{}
The formulae reviewed above, except for that regarding the transition 
frequencies (Eqs.~(\ref{eq7}) and (\ref{eq8})), are classical, and almost the same ones have 
previously been used in Ref.~\cite{ref4}. As will be shown in the next section, 
however, the following investigation based on Eq.~(\ref{eq7}) coupled with Eq.~(\ref{eq9}) 
gives a different interpretation of the sign reversal from the previous ones 
described using only the natural frequencies.

\section{RESULTS AND DISCUSSION}
In this section, we investigate the relationship between the transition 
frequencies and the sign of the secondary Bjerknes force by using some 
examples. The first example is the case of $R_{10} = 2$ mm and $R_{20} = 5$ mm, which corresponds to a case used in Ref.~\cite{ref9}. We 
assume that the bubbles are filled with a gas having a specific heat ratio 
of $\gamma  = 1.4$, and the surrounding material is water ($\sigma  = 0.0728$ N/m, $\rho  = 1000$ kg/m$^3$, $P_0  = 1$ atm, and the speed of sound $c = 1500$ m/s). For the damping coefficient, we adopt that used for radiation and thermal 
losses:
\begin{equation}
\label{eq10}
\delta _j  = \frac{{\omega ^2 R_{j0} }}{c} + \beta _{{\rm th},j} ,
\end{equation}
where the thermal damping coefficient $\beta _{{\rm th},j}$ and the effective polytropic exponent $\kappa _j$ are determined by \cite{ref21,ref15,ref8}
\begin{eqnarray*}
&&\beta _{{\rm th},j}  = \frac{{\omega _{j0}^2 }}{\omega }d_{{\rm th},j} , \\
&&\kappa _j  = \gamma \left[ {\left( {1 + d_{{\rm th},j}^2 } \right)\left( {1 + \frac{{3(\gamma  - 1)(\sinh X - \sin X)}}{{X(\cosh X - \cos X)}}} \right)} \right]^{ - 1} ,
\end{eqnarray*}
with
\begin{eqnarray*}
&&d_{{\rm th},j}  = 3(\gamma  - 1) \\ &&\times 
\frac{{X(\sinh X + \sin X) - 2(\cosh X - \cos X)}}{{X^2 (\cosh X - \cos X) + 3(\gamma  - 1)X(\sinh X - \sin X)}} , \\
&&X = R_{j0} (2\omega /\chi _G )^{1/2} ,
\end{eqnarray*}
where we set $\chi _G  = 2 \times 10^{ - 5}$ m$^2$ s$^{- 1}$.

\begin{figure}
\begin{center}
\includegraphics[width=8.8cm]{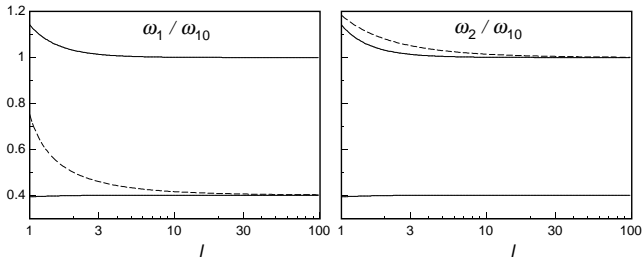}
\caption{Transition frequencies $\omega _{1}$ (rad/s) and $\omega _{2}$ (rad/s) for $R_{10} = 2$ mm, $R_{20} = 5$ mm, and the reduced damping, normalized by $\omega _{10}$ (rad/s). The dashed lines show the transition frequencies that do not cause the resonance.}
\label{fig1}
\end{center}
\end{figure}

\begin{figure}
\begin{center}
\includegraphics[width=8.8cm]{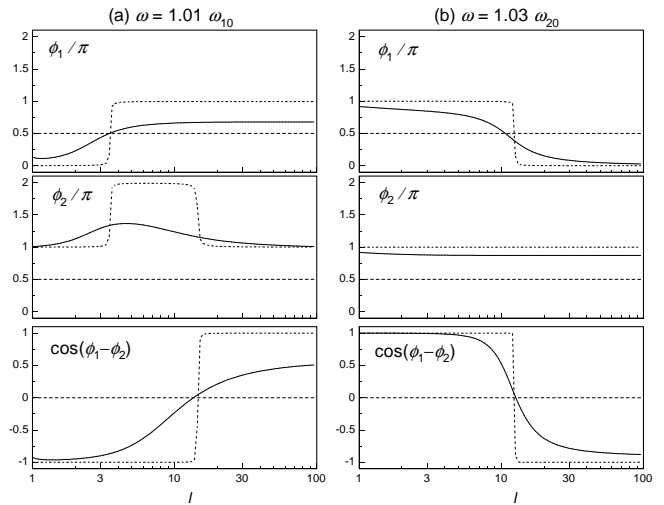}
\caption{$\phi _1 / \pi$, $\phi _2 / \pi$, and $\cos (\phi _1 - \phi _2 )$ for (a) $\omega = 1.01\omega _{10}$ (rad/s) and (b) $\omega = 1.03\omega _{20}$ (rad/s), as functions of $l$. The dashed curves and the solid lines show the results given using the reduced damping and the real damping, respectively.}
\label{fig2}
\end{center}
\end{figure}

In order to clarify the following discussion, we first present results for 
the idealized condition of $\delta _j \approx 0$ by resetting $\delta _j \to \delta _j /100$, and subsequently provide results given by the direct use of Eq.~(\ref{eq10}). Figure \ref{fig1} shows the transition frequencies of the bubbles, 
$\omega _1$ and $\omega _2$, calculated using 
Eq.~(\ref{eq7}) with the reduced damping, normalized by $\omega _{10}$ ($ = \omega _1$ for $D \to \infty$). In those figures, $l$ denotes the normalized distance defined as $l = D/(R_{10}  + R_{20} )$. As mentioned previously, we can 
observe three transition frequencies, only one of which converges to 
$\omega _{j0}$ of the corresponding bubble for $l \to \infty$. The second-highest transition frequency of bubble 2 is almost 
equal to the highest one of bubble 1; thus, the highest one of bubble 2 is 
higher than that of bubble 1. The second highest one of bubble 1 and the 
highest one of bubble 2 do not cause the resonance response \cite{ref14}.

The dashed curves displayed in Fig.~\ref{fig2}(a) show $\phi _1$, $\phi _2$, and $\cos (\phi _1  - \phi _2 )$, respectively, as functions of $l$. Here 
the driving frequency is assumed to be $\omega  = 1.01\omega _{10}$, i.e., slightly above $\omega _{10}$. (In the present study, the driving frequency is 
set as $\omega  \approx \omega _{10}$ or $\omega  \approx \omega _{20}$ so that the sign reversal takes place at a sufficiently large $l$ where the accuracy of Eqs.~(\ref{eq5}) and (\ref{eq6}) is 
guaranteed.) 
As mentioned in Sec.~\ref{sec1}, it is known already that the sign 
reversal can take place when $\omega  > \omega _{10}  > \omega _{20}$ or $\omega _{10}  > \omega  > \omega _{20}$; the 
present setting corresponds to the former case. We can observe one and two 
sharp shifts of $\phi _1$ and $\phi _2$, respectively. At $l \approx 3$, both 
$\phi _1$ and $\phi _2$ shift almost 
simultaneously, but the sign reversal does not occur because the phase 
difference $\phi _1  - \phi _2$ is hardly changed. At $l \approx 13$, only $\phi _2$ shifts, resulting in the sign reversal. In the former case, the phase shifts are 
caused by the natural frequencies. As mentioned previously, when 
$\delta _j  \approx 0$ both the bubbles 
have (almost) the same natural frequencies. The simultaneous phase shift, 
thus, appears. The change of $\phi _2 $ in the later case is apparently due to the highest transition 
frequency of bubble 2, which cannot be obtained by the traditional 
natural-frequency analysis. Namely, this sign reversal cannot be interpreted 
by using only the natural frequencies.

We should note here that, to compute the phase delays $\phi _1$ and $\phi _2$, we used the ``${\rm atan2}(a, b)$'' function in the C language, which returns $\tan ^{ - 1} (b/a) \in [- \pi, \pi]$, and, furthermore, adopted the operation
\[
\phi _j  = \left\{ {\begin{array}{*{20}c}
   {{\rm atan2}(A_j, B_j ) + 2\pi} \hfill & {{\rm if}\,\,\, {\rm atan2}(A_j, B_j ) < 0,} \hfill  \\
   {{\rm atan2}(A_j ,B_j )} \hfill & {{\rm otherwise}} \hfill  \\
\end{array}} \right.
\]
in order to obtain results for $l \gg 1$ that are consistent with the established knowledge of single-bubble dynamics, e.g., $\phi _j \approx \pi$ when $\omega  > \omega _{j0}$ and $\delta _j \approx 0$ ($j = 1$ or $2$).

The dashed curves displayed in Fig.~\ref{fig2}(b) show results for $\omega  = 1.03\omega _{20}$ ($ = 0.413\omega _{10}$), i.e., 
for $\omega _{10}  > \omega  > \omega _{20}$. In this case, 
we can observe only one sharp shift of $\phi _1$ at $l \approx 12$, causing the sign reversal. This shift of $\phi _1$ is due to the second-highest 
transition frequency of bubble 1 (this frequency also not corresponding to 
the natural frequency!), because the lowest transition frequencies of both 
the bubbles decrease as $l$ decreases.

These results reveal that in the above cases, the transition frequencies 
other than the natural frequencies cause the sign reversal of the secondary 
Bjerknes force. This conclusion is obviously different from the previous 
interpretations described by means of the natural frequencies \cite{ref4,ref8,ref9,ref13}.

It is interesting to point out that, in the case where $\omega  > \omega _{10}  > \omega _{20}$ and $\omega  \approx \omega _{10}$, the phase delay of the larger bubble was sometimes greater than $\pi$ (see Fig.~\ref{fig2}(a)). Such a result cannot be given by a single-bubble model which 
predicts a phase delay of up to $\pi$. This may be explained as follows: When 
$\omega  > \omega _{10}  > \omega _{20}$ is true and 
$l$ is sufficiently 
large, both bubbles pulsate out-of-phase with $p_{{\rm ex}}$, emitting sound waves whose phases are also out-of-phase with $p_{{\rm ex}}$. 
As $l$ decreases, if $\omega  \approx \omega _{10}$, the amplitude of 
the sound wave emitted by bubble 1 measured at ${\bf r}_2$ can be greater than the amplitude of $p_{{\rm ex}} $. In this situation, 
bubble 2 is driven by a sound wave whose oscillation phase is delayed by 
almost $\pi$ from that of $p_{{\rm ex}} $. This results in 
$\phi _2  > \pi $, because the 
pulsation phase of bubble 2 delays further from that of the sound wave.

\begin{figure}
\begin{center}
\includegraphics[width=8.8cm]{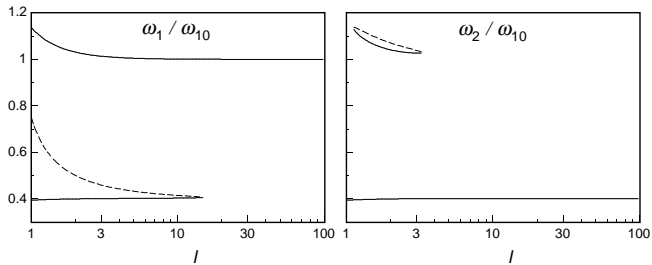}
\caption{Same as in Fig.~\ref{fig1}, but for the real damping.}
\label{fig3}
\end{center}
\end{figure}

We show the here results given by using Eq.~(\ref{eq10}) in order to examine 
the influences of the damping effects on the sign reversal and phase shifts. 
Figure \ref{fig3} shows the recalculated transition frequencies. As already discussed \cite{ref14}, when the damping effects are not negligible, the bubbles have only one 
transition frequency in the large-$l$ region. The solid curves displayed in Fig.~\ref{fig2} show $\phi _1$, $\phi _2$, and $\cos (\phi _1  - \phi _2 )$ for $\omega  = 1.01\omega _{10}$ and $\omega  = 1.03\omega _{20}$. Their tendencies 
are similar to those given with the reduced damping, although their profiles 
are smoothed significantly (Such a smoothing of the phase change by the 
damping effects is well known for a single-bubble case) and the points at 
which the sign reversal takes place are shifted slightly; the positions of 
these points are, in the case of $\omega  = 1.01\omega _{10}$, $l \approx 13.59$ for $\delta _j $ and $l \approx 14.57$ for $\delta _j /100$, and, in the case of $\omega  = 1.03\omega _{20} $, 
$l \approx 12.66$ for $\delta _j $ and $l \approx 12.48$ for $\delta _j /100$. Moreover, $\phi _2 $ for $\omega  = 1.01\omega _{10} $ does not exceed 
$3\pi /2$ (the minimum value of $\omega _2 $ larger than $\omega _{10} $ is 
${\rm 1}{\rm .027}\omega _{10}$.); even so, the 
sign reversal occurs at almost the same point as that given with 
$\delta _j /100$, away from the point where $\phi _1  = \pi /2$. This 
result may be interpreted as the ``vestige'' of the highest transition 
frequency of the larger bubble having given rise to this sign reversal. 
Detailed theoretical discussions for the slight shift in $l$ for $\cos (\phi _1  - \phi _2 ) = 0$ due to the damping effects will be provided in a 
future paper.

\begin{figure}
\begin{center}
\includegraphics[width=8.8cm]{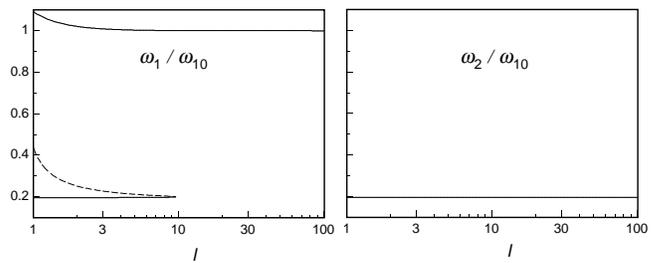}
\caption{Transition frequencies $\omega _{1}$ (rad/s) and $\omega _{2}$ (rad/s) for $R_{10} = 1$ $\mu$m, $R_{20} = 4$ $\mu$m, and the real damping, normalized by $\omega _{10}$ (rad/s).}
\label{fig4}
\end{center}
\end{figure}

\begin{figure}
\begin{center}
\includegraphics[width=8.8cm]{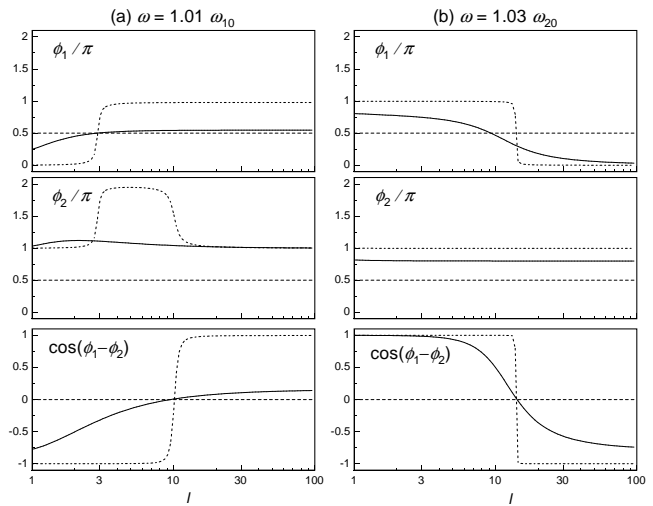}
\caption{$\phi _1 / \pi$, $\phi _2 / \pi$, and $\cos (\phi _1 - \phi _2 )$ for (a) $\omega = 1.01\omega _{10}$ (rad/s) and (b) $\omega = 1.03\omega _{20}$ (rad/s). The dashed curves denote the results for the reduced damping.}
\label{fig5}
\end{center}
\end{figure}

\begin{figure}
\begin{center}
\includegraphics[width=8.8cm]{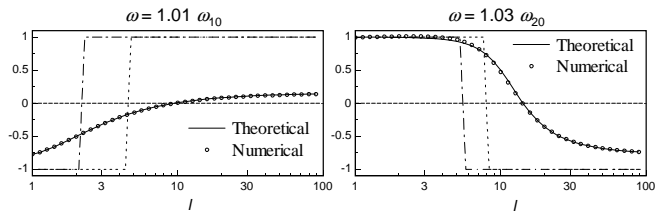}
\caption{Comparison between the theoretical and the numerical results. The 
lines and the circles denote the theoretical and the numerical results, 
respectively, of $\cos (\phi _1 - \phi _2 )$ for $\omega = 1.01\omega _{10}$ (rad/s) and $1.03\omega _{20}$ (rad/s). Additionally, ${\mathop{\rm sgn}} \left( \left\langle {\dot V_1 \dot V_2 } \right\rangle \right)$ for $P_a / P_0 = 0.2 $ (the dots) and $0.5$ (the dash-dotted curves) are plotted for a brief investigation of nonlinear effects.}
\label{fig6}
\end{center}
\end{figure}

Next, we show results for smaller bubbles ($R_{10}  = 1$ $\mu$m and $R_{20}  = 4$ $\mu$m). The value for viscous loss is used for the 
damping coefficients, i.e.,
\begin{equation}
\label{eq11}
\delta _j  = \frac{{4\mu }}{{\rho R_{j0}^2 }} ,
\end{equation}
where the viscosity of water $\mu  = 1.002 \times 10^{ - 3}$ kg/(m s). Because the thermal effect is neglected, $\kappa  = \gamma  = 1.4$. Figure \ref{fig4} shows the 
transition frequencies, and Figure \ref{fig5} shows $\phi _1 $, $\phi _2 $, and $\cos (\phi _1  - \phi _2 )$ for $\omega  = 1.01\omega _{10}$ and $\omega  = 1.03\omega _{20}$ ($ = 0.201\omega _{10}$) with $\delta _j /100$ (the dashed curves) and $\delta _j$ (the solid curves). The qualitative 
natures of these results are quite similar with the previous ones; thus, 
additional discussion may not be necessary. Using this example, we perform 
here a comparative study of the theoretical results with the numerical 
results in order to confirm the former's correctness. In the numerical 
experiment, we employ the coupled RPNNP (Rayleigh, Plesset, Noltingk, Neppiras, 
and Poritsky) equations (see, e.g., Ref.~\cite{ref14});
\begin{eqnarray*}
R_1 \ddot R_1  + \frac{3}{2}\dot R_1^2  - \frac{1}{\rho }p_{w,1}  =  - \frac{1}{\rho }\left[ {p_{{\rm ex}}  + \frac{\rho }{D}\frac{d}{{dt}}(R_2^2 \dot R_2 )} \right] , \\
R_2 \ddot R_2  + \frac{3}{2}\dot R_2^2  - \frac{1}{\rho }p_{w,2}  =  - \frac{1}{\rho }\left[ {p_{{\rm ex}}  + \frac{\rho }{D}\frac{d}{{dt}}(R_1^2 \dot R_1 )} \right] ,
\end{eqnarray*}
where

\begin{figure}
\begin{center}
\includegraphics[width=8.6cm]{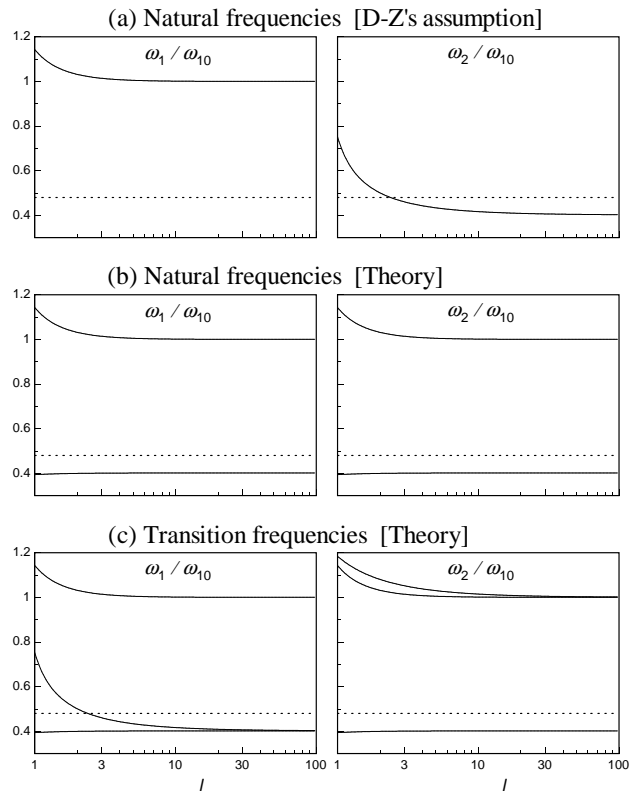}
\caption{Characteristic frequencies of two coupled bubbles and different 
interpretations of the sign reversal. The dashed lines show a typical 
driving frequency lying between $\omega _{10}$ and $\omega _{20} $, where $\omega _{10}  > \omega _{20} $ is 
assumed. Doinikov and Zavtrak assumed that the natural frequencies of both 
the bubbles increase as the bubbles approach each other (a). Assuming this, 
the sign reversal for $\omega _{10}  > \omega  > \omega _{20}$ seems to be explained. This assumption is, however, inconsistent 
with the classical theory for natural frequencies (b). The present theory 
can explain this reversal without such an inconsistency (c).}
\label{fig7}
\end{center}
\end{figure}

\[
p_{w,j}  = \left( {P_0  + \frac{{2\sigma }}{{R_{j0} }}} \right)\left( {\frac{{R_{j0} }}{{R_j }}} \right)^{3\kappa }  - \frac{{2\sigma }}{{R_j }} - \frac{{4\mu \dot R_j }}{{R_j }} - P_0 .
\]
This system of nonlinear differential equations are solved numerically 
through the use of the fourth-order Runge-Kutta method in which 
$R_1 $, $R_2 $, $\dot R_1 $, and $\dot R_2 $ are used as dependent variables, and $\left\langle {R_1^2 \dot R_1 R_2^2 \dot R_2 } \right\rangle$ [$ \propto \left\langle {\dot V_1 \dot V_2 } \right\rangle$ in Eq.~(\ref{eq9})] is then calculated. The time average is performed 
during a sufficiently large period after the transients have decayed. 
Normalizing $\left\langle {R_1^2 \dot R_1 R_2^2 \dot R_2 } \right\rangle$ by 
$R_{10}^2 R_{20}^2 \max (\left| {R_1 (t) - R_{10} } \right|)\max (\left| {R_2 (t) - R_{20} } \right|)\omega ^2 /2$ yields the 
numerical approximation of $\cos (\phi _1  - \phi _2 )$, where $\max (\left| {R_j (t) - R_{j0} } \right|)$ 
indicates the pulsation amplitude of bubble $j$ given numerically. The amplitude of the external sound is set to $P_a  = 0.01P_0 $. In Figure \ref{fig6}, the numerical and the theoretical results are displayed in piles. 
These results are in excellent agreement, confirming the correctness of the theoretical results given above. 
In the same figure, we have shown additionally ${\mathop{\rm sgn}} \left(\left\langle {\dot V_1 \dot V_2 } \right\rangle \right)$ for $P_a = 0.2P_0$ (the dots) and $0.5P_0$ (the dash-dotted curves) in order to briefly investigate nonlinear effects on the sign reversal, where ${\mathop{\rm sgn}} (X) = 1$ for $X > 0$ and ${\mathop{\rm sgn}} (X) = - 1$ otherwise. 
In plotting these results, we omitted the data in the case where $R_1 (t) + R_2 (t) > D$ was observed during the computation. 
As is clearly shown, increasing the driving pressure reduces the distance for which the sign reversal takes place. This result appears to be consistent with the well-known nonlinear phenomenon that a strong driving pressure decreases a bubble's (effective) resonance frequency, see, e.g., Refs.~\cite{ref2,ref3}. (Imagine that the transition frequencies shown, e.g., in Fig.~\ref{fig1} decrease but the driving frequency holds, which might shorten the distance for ${\bf F} = 0$.) More detailed and concrete discussions on the nonlinear effects will be provided in a future paper.

To summarize our discussion, we compare the present interpretation with the 
previous ones. Figure \ref{fig7}(a) shows the dependency of natural frequencies on 
$l$ that Doinikov and Zavtrak assumed \cite{ref8,ref9}. Their assumption explains the sign reversal 
occurring when $\omega _{10}  > \omega  > \omega _{20}$, for 
example, as taking place around $l$ at which $\omega _2 = \omega $ is true. Yet, as mentioned, their assumption is 
inconsistent with the theoretical results regarding natural frequencies 
given previously \cite{ref4,ref5} (see Fig.~\ref{fig7}(b)). On the other hand, it is difficult 
to determine by only observing the natural frequencies that the sign 
reversal can take place for $\omega _{10}  > \omega  > \omega _{20}$, because the classical theory does not show that a kind of 
characteristic frequency exists in the frequency region between $\omega _{10}$ and $\omega _{20}$. The present theory explains the sign reversal in this case as taking place around $l$ at which $\omega _1  = \omega $ is true (see Fig.~\ref{fig7}(c), where we assume for simplicity that the damping effect is negligible), and is consistent with the theory for natural frequencies because the transition frequencies include the natural frequencies.

\section{CONCLUSION}
We have investigated the influences of change in the transition frequencies 
of gas bubbles, resulting from their radiative interaction, on the sign of 
the secondary Bjerknes force. The most important point suggested in this 
paper is that the transition frequencies that cannot be derived by the 
natural-frequency analysis cause the sign reversal in the cases of both 
$\omega  > \omega _{10}  > \omega _{20}$ and $\omega _{10}  > \omega  > \omega _{20}$. This 
interpretation has not been proposed previously. The present results also 
show that the theory given in Ref.~\cite{ref14} for evaluating the transition 
frequencies of interacting bubbles is a reasonable tool for accurately 
understanding the mechanism of this reversal. In a paper currently in 
preparation \cite{ref16}, we will use the direct numerical simulation technique \cite{ref17,ref18} to verify the present theoretical results.

Lastly, we make further remarks regarding the results described in Ref.~\cite{ref9}. 
In that paper, the frequency of the external sound ($f = \omega /2\pi$) was assumed to be $f = 63$ kHz, which is 60 times higher than the 
partial resonance frequency of a bubble of $R_0  = 3$ mm (1.094 kHz); nevertheless, the reversal was 
observed at a very small $l$. (In Ref.~\cite{ref8}, the driving frequency is assumed to be comparable 
to the partial natural frequencies of bubbles, and the bubble radii are 
several tens of micrometers.) This result reveals implicitly that the 
mathematical model proposed in Ref.~\cite{ref8}, which takes into account the shape 
deviation of the bubbles, predicts such a strong increase of the transition 
frequencies of closely coupled large bubbles that this increase cannot be 
explained by the classical model for coupled oscillators used here. 
Derivation of the transition frequencies of Doinikov and Zavtrak's model 
would be an interesting subject for future study.

\begin{acknowledgments}
This work was supported by the Ministry of Education, Culture, Sports, 
Science, and Technology of Japan (Monbu-Kagaku-Sho) under an IT research 
program "Frontier Simulation Software for Industrial Science."
\end{acknowledgments}


\begin{thebibliography}{}
\bibitem{ref22} Strictly speaking, the natural frequency is different from the 
resonance frequency. Because these frequencies are almost equal to each 
other in many cases, however, we treat them as the same, as has been done 
frequently.
\end{thebibliography}
\end{document}